\documentclass[11pt,letter]{article}
\usepackage[papersize={8.5in,11in},margin=1in]{geometry}
\usepackage{amssymb}
\usepackage{amsmath}
\usepackage{booktabs}
\usepackage{graphicx}
\usepackage{makeidx}
\usepackage{multicol}
\usepackage{algorithm}
\usepackage{algorithmic}
\usepackage{subcaption}
\usepackage[english]{babel}
\usepackage{color}

\usepackage{flexisym}
\usepackage{authblk}
\usepackage{tikz}
\usetikzlibrary{shapes.geometric, arrows}

\tikzstyle{process} = [rectangle, rounded corners, minimum width=2.5cm, minimum height=3cm,text centered, text width=2.5cm, draw=black, fill=red!30]
\tikzstyle{process_top} = [rectangle, rounded corners, minimum width=2.5cm, minimum height=3cm,text centered, text width=2.5cm, draw=black, fill=red!30]
\tikzstyle{io} = [trapezium, trapezium left angle=70, trapezium right angle=110, minimum width=3cm, minimum height=1cm, text centered, draw=black, fill=blue!30]
\tikzstyle{textbox} = [rectangle, minimum width=3cm, minimum height=1cm, text centered, text width=3cm, draw=black, fill=orange!30]
\tikzstyle{decision} = [diamond, minimum width=3cm, minimum height=1cm, text centered, draw=black, fill=green!30]
\tikzstyle{arrow} = [thick,->,>=stealth]
\tikzstyle{dashed_arrow} = [thick,dashed,->,>=stealth]

\newcommand{\walklets}{\textsc{Walklets}}

\newtheorem{definition}{Definition}

\newcommand{\todo}[1]{}
\renewcommand{\todo}[1]{{\color{red} TODO: {#1}}}

\begin{document}
\title{A Tutorial on Network Embeddings}

\author[1]{Haochen Chen}
\author[2]{Bryan Perozzi}
\author[2]{Rami Al-Rfou}
\author[1]{Steven Skiena}
\affil[1]{Stony Brook University}
\affil[2]{Google Research}
\affil[ ]{\texttt{\{haocchen, skiena\}@cs.stonybrook.edu, bperozzi@acm.org, rmyeid@google.com}}

\maketitle

\begin{abstract}
Network embedding methods aim at learning low-dimensional latent representation of nodes in a network.
These representations can be used as features for a wide range of tasks on graphs such as classification,
clustering, link prediction, and visualization.
In this survey, we give an overview of network embeddings by summarizing and categorizing recent advancements in this research field.
We first discuss the desirable properties of network embeddings and
briefly introduce the history of network embedding algorithms.
Then, we discuss network embedding methods under different scenarios, such as supervised versus unsupervised learning,
learning embeddings for homogeneous networks versus for heterogeneous networks, etc.
We further demonstrate the applications of network embeddings,
and conclude the survey with future work in this area.
\end{abstract}

\section{Introduction}\label{intro}
From social networks to the World Wide Web, networks provide a ubiquitous way to organize a diverse set of real-world information.
Given a network's structure, it is often desirable to predict missing information (frequently called \textit{attributes} or \textit{labels}) associated with each node in the graph.
This missing information can represent a variety of aspects of the data -- for example, on a social network they might represent the communities a person belongs to, or the categories of a document's content on the web.

Because information networks can contain billions of nodes and edges, it can be intractable to perform complex inference procedures on the entire network.
One technique which has been proposed to address this problem is \textit{network embedding}.
The central idea is to find a mapping function which converts each node in the network to a low-dimensional latent representation.
These representations can then be used as features for common tasks on graphs such as classification, clustering, link prediction, and visualization.

To sum up, we seek to learn network embeddings with the following characteristics:
\begin{itemize}
\item \textbf{Adaptability} - Real networks are constantly evolving;  new applications should not require repeating the learning process all over again.

\item \textbf{Scalability} - Real networks are often large in nature, thus network embedding algorithms should be able to process large-scale networks in a short time period.

\item \textbf{Community aware} - The distance between latent representations should represent a metric for evaluating similarity between the corresponding members of the network.
This allows generalization in networks with homophily.

\item \textbf{Low dimensional} - When labeled data is scarce, low-dimensional models generalize better, and speed up convergence and inference.

\item \textbf{Continuous} -
We require latent representations to model partial community membership in continuous space.
In addition to providing a nuanced view of community membership, a continuous representation has smooth decision boundaries between communities which allows more robust classification.
\end{itemize}

\begin{figure}[htb]
	\centering
        \begin{subfigure}[b]{0.54\textwidth}
                \includegraphics[width=\textwidth]{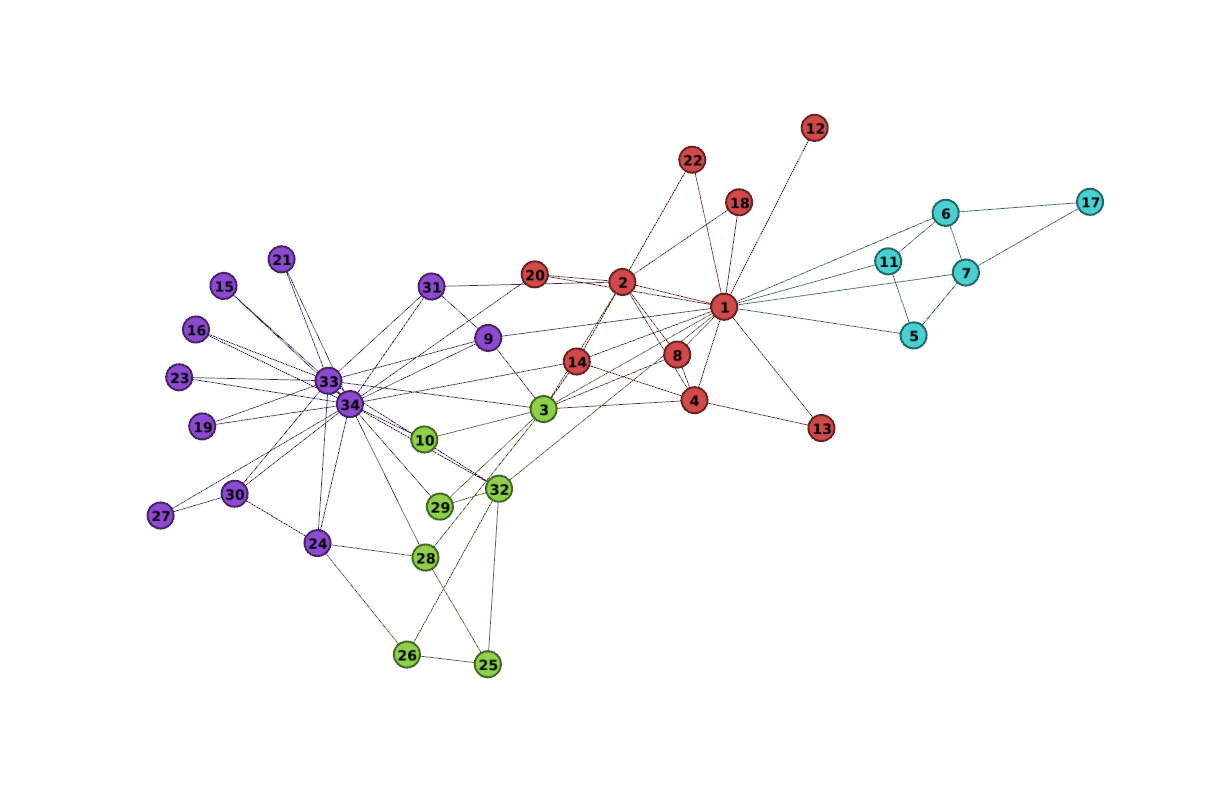}
                \caption{Input: Karate Graph}
                \label{fig:toy_example_graph}
        \end{subfigure}
        \begin{subfigure}[b]{0.44\textwidth}
                \includegraphics[width=1.1\textwidth]{figures/karate}
                \caption{Output: Network Embedding}
                \label{fig:toy_example_embedding}
        \end{subfigure}
        \caption{Network embedding methods \emph{learn} latent representation of nodes in a network in $\mathbb{R}^d$.  This learned representation encodes community structure so it can be easily exploited by standard classification methods. Here, DeepWalk \cite{deepwalk} is used on Zachary's Karate network \cite{zachary1977information} to generate a latent representation in $\mathbb{R}^2$.
		Note the correspondence between community structure in the input graph and the embedding. Vertex colors represent a modularity-based clustering of the input graph.
        }
\label{fig:toy_example}
\end{figure}

As a motivating example we show the result of applying DeepWalk \cite{deepwalk}, which is a widely used network embedding method, to the well-studied Karate network in Figure \ref{fig:toy_example}.
This network, as typically presented by force-directed layouts, is shown in Figure \ref{fig:toy_example}.
Figure \ref{fig:toy_example_embedding} shows the output of DeepWalk with two latent dimensions.
Beyond the striking similarity, note that linearly separable portions of (\ref{fig:toy_example_embedding}) correspond to clusters found through modularity maximization in the input graph (\ref{fig:toy_example_graph}) (shown as vertex colors).

The rest of the survey is organized as follows.\footnote{The area of network embedding is rapidly growing, and while we have made an effort to include all relevant work in this survey, there have doubtlessly been accidental omissions.  If you are aware of work that would improve the completeness of this survey, please let the authors know.}
We first provide a general overview of network embedding and give some definitions and notations which will be used later.
In Section \ref{sec:unsupervised_network_embedding}, we introduce unsupervised network embedding methods on homogeneous networks without attributes.
Section \ref{sec:attributed_network_embedding} reviews embedding methods on attributed networks and partially labeled networks.
Then, Section \ref{sec:heterogeneous_network_embedding} discusses heterogeneous network embedding algorithms.
We further demonstrate the applications of network embeddings,
and conclude the survey with future work in this area.

\subsection{A Brief History of Network Embedding}
Traditionally, graph embeddings have been described in the context of
{\em dimensionality reduction}.
Classical techniques for dimensionality reduction include principal component analysis (PCA) \cite{wold1987principal}
and multidimensional scaling (MDS) \cite{kruskal1978multidimensional}.
Both methods seek to represent an $n \times m$ matrix $M$
as a $n \times k$ matrix where $k << n$.
For graphs, $M$ is typically an $n \times n$ matrix, where could be the adjacency matrix, normalized Laplacian matrix or all-pairs shortest path matrix, to name a few.
Both methods are capable of capturing linear structural information,
but fails to discover the non-linearity within the input data.

\textbf{PCA - } PCA computes a set of orthogonal \emph{principal components}, where each principal components is a linear combinations of the original variables.
The number of these components could be equal or less than $m$, which is the reason that PCA can serve as a dimensionality reduction technique.
Once the principal components are computed, each original data point could be projected to the lower-dimensional space determined by them.
For a square matrix, the time complexity of PCA is $O(n^3)$.

\textbf{MDS - } Multidimensional scaling (MDS) projects each row of $M$ to a $k$-dimensional vector,
such that the distance between different objects in the original feature matrix $M$ is best preserved in the $k$-dimensional space.
Specifically, let $y_i \in \mathbb{R}^k$ to be the coordinate of the $i$-th object in the embedding space, metric MDS minimizes the following \textit{stress} function:
\begin{equation}
	Stress(y_1, y_2, \cdots, y_n) = \left( \sum_{i, j = 1, 2, \cdots, n}(M_{ij} - \lVert y_i - y_j \rVert)^2 \right)^{1/2}
\end{equation}
Exact MDS computation requires eigendecomposition of a transformation of $M$, which takes $O(n^3)$ time.

In early 2000s, other methods such as IsoMap \cite{isomap} and locally linear embeddings (LLE) \cite{lle} were proposed to preserve the global structure of non-linear manifolds.
We note that both methods are defined abstractly for any type of dataset, and first preprocesses the data points into graphs which capture local neighborhood performance.

\textbf{Isomap - } Isomap \cite{isomap} is an extension to MDS with the goal of preserving geodesic distances in the neighborhood graph of input data. The neighborhood graph $G$ is constructed by connecting each node $i$ with either nodes closer than a certain distance $\epsilon$ or
nodes which are $k$-nearest neighbors of $i$. Then, classical MDS is applied to $G$ to map data points to a low-dimensional manifold which preserves geodesic distances in $G$.

\textbf{Local Linear Embeddings (LLE) - } Unlike MDS, which preserves pairwise distances between feature vectors, LLE \cite{lle} only exploits the local neighborhood of data points and does not attempt to estimate distance between distant data points.
LLE assumes that the input data is intrinsically sampled from a latent manifold, and that a data point can be reconstructed from a linear combination of its neighbors. The reconstruction error can be defined as
\begin{equation}
	E(W) = \sum_{i} |x_i - \sum_{j}W_{ij}x_j|^2
\end{equation}
where $W$ is a weight matrix denoting data point $j$'s contribution to $i$'s reconstruction,
which is computed by minimizing the loss function above.
Since $W_{ij}$ reflects the invariant geometric properties of the input data,
it can used to find the mapping from a data point $x_i$ to its low-dimensional representation $y_i$.
To compute $y_i$, LLE minimizes the following embedding cost function:
\begin{equation}
	\Phi(Y) = \sum_{i} |y_i - \sum_{j}W_{ij}y_j|^2
\end{equation}
Since $W_{ij}$ is already fixed,
it can be proved that this cost function can be minimized by finding the eigenvectors of an auxiliary matrix.

In general, these methods all offer good performance on small networks.
However, the time complexity of these methods are at least quadratic,
which makes them impossible to run on large-scale networks.

Another popular class of dimensionality reduction techniques uses the spectral properties (e.g.\ eigenvectors) of matrices derivable from the graph (e.g.\ Graph Laplacian) to embed the nodes of the graph.
\textbf{Laplacian eigenmaps (LE)} \cite{belkin2002laplacian}, represent each node in the graph by the eigenvectors associated with its $k$-smallest nontrivial eigenvalues.
The spectral properties of the Graph Laplacian encode cut information about the graph, and have a rich history of use in graph analysis \cite{chung1997spectral}.
Let $W_{ij}$ be the weight of the connection between node $i$ and $j$, the diagonal weight matrix $D$ can be constructed:
\begin{equation}
D_{ii} = \sum_{j}W_{ji}
\end{equation}
The Laplacian matrix of $M$ is then
\begin{equation}
L = D - W
\end{equation}
The solutions to the eigenvector problem:
\begin{equation}
L\mathbf{f} = \lambda D\mathbf{f}
\end{equation}
can be used as the low-dimension embeddings of the input graph.

Tang and Liu \cite{tang2011leveraging} examined using eigenvectors of the Graph Laplacian for classification in social networks.
They argue that  nodes (actors) in a network are associated with different latent affiliations.
On the other hand, these social dimensions should also be continuous since the actors might have different magnitude of associations to one affiliation.
Another similar method,  SocDim \cite{tang2009relational} proposed using the spectral properties of the modularity matrix as latent social dimensions in networks.
However, the performance of these methods has been shown to be lower than neural network-based approaches \cite{deepwalk}, which we will shortly discuss.

\begin{figure}[t!]
	\centering
	\begin{subfigure}[b]{0.32\textwidth}
		\includegraphics[width=\textwidth]{figures/karate}
		\caption{Output: DeepWalk}
	\end{subfigure}
	\begin{subfigure}[b]{0.32\textwidth}
		\includegraphics[width=\textwidth]{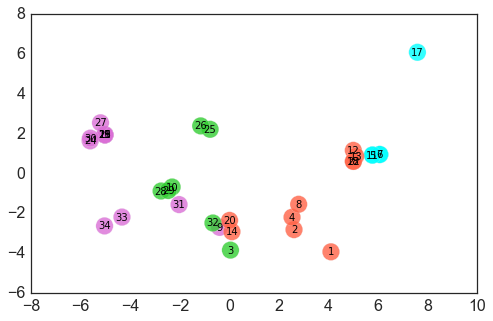}
		\caption{Output: PCA}
	\end{subfigure}
	\begin{subfigure}[b]{0.32\textwidth}
		\includegraphics[width=\textwidth]{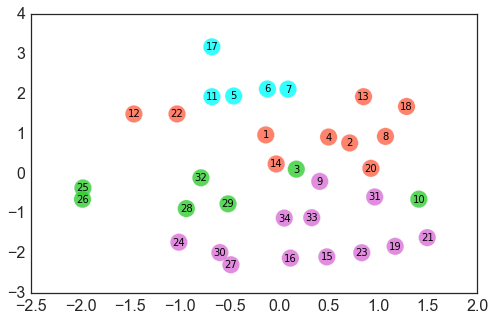}
		\caption{Output: MDS}
	\end{subfigure}

	\begin{subfigure}[b]{0.32\textwidth}
		\includegraphics[width=\textwidth]{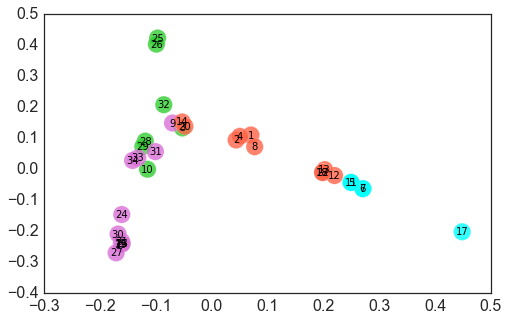}
		\caption{Output: LLE}
	\end{subfigure}
	\begin{subfigure}[b]{0.32\textwidth}
		\includegraphics[width=\textwidth]{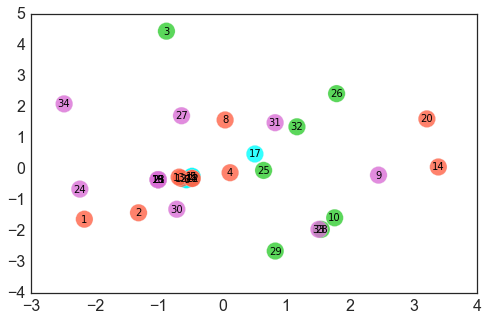}
		\caption{Output: LE}
	\end{subfigure}
	\begin{subfigure}[b]{0.32\textwidth}
		\includegraphics[width=\textwidth]{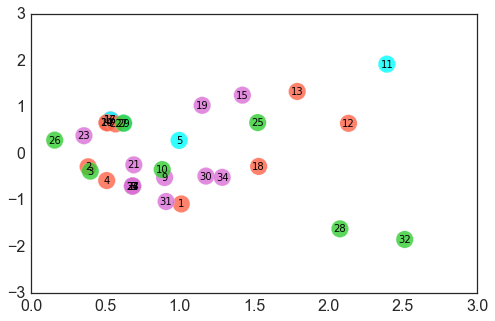}
		\caption{Output: SVD}
	\end{subfigure}
	\caption{
		Two-dimensional embeddings of the Karate graph using DeepWalk and several early dimension reduction techniques.
		The input is the adjacency matrix for DeepWalk and SVD, and the geodesic matrix for the other four methods.
	}
	\label{fig:toy_example_2}
\end{figure}

\subsection{The Age of Deep Learning}
DeepWalk \cite{deepwalk} was proposed as the first network embedding method using techniques from the representation learning \emph{(or deep learning)} community.
DeepWalk bridges the gap between network embeddings and word embeddings by treating nodes as words and generating short random walks as sentences.
Then, neural language models such as Skip-gram \cite{skipgram} can be applied on these random walks to obtain network embedding.
DeepWalk has become arguably the most popular network embedding method since then, for several reasons.

\begin{figure}[t!]
\centering
\begin{tikzpicture}[node distance=2cm]
\node (input) [process] {\textbf{Input}: Choose a matrix associated with the input graph};
\node (graph_sampling) [process, right of=input, xshift=2cm] {\textbf{Graph Sampling}: Sample sequences from the chosen matrix};
\node (modeling) [process, right of=graph_sampling, xshift=2cm] {\textbf{Modeling}: Learn embeddings from the sequences or the matrix itself};
\node (output) [process_top, right of=modeling, xshift=2cm] {\includegraphics[width=2.4cm]{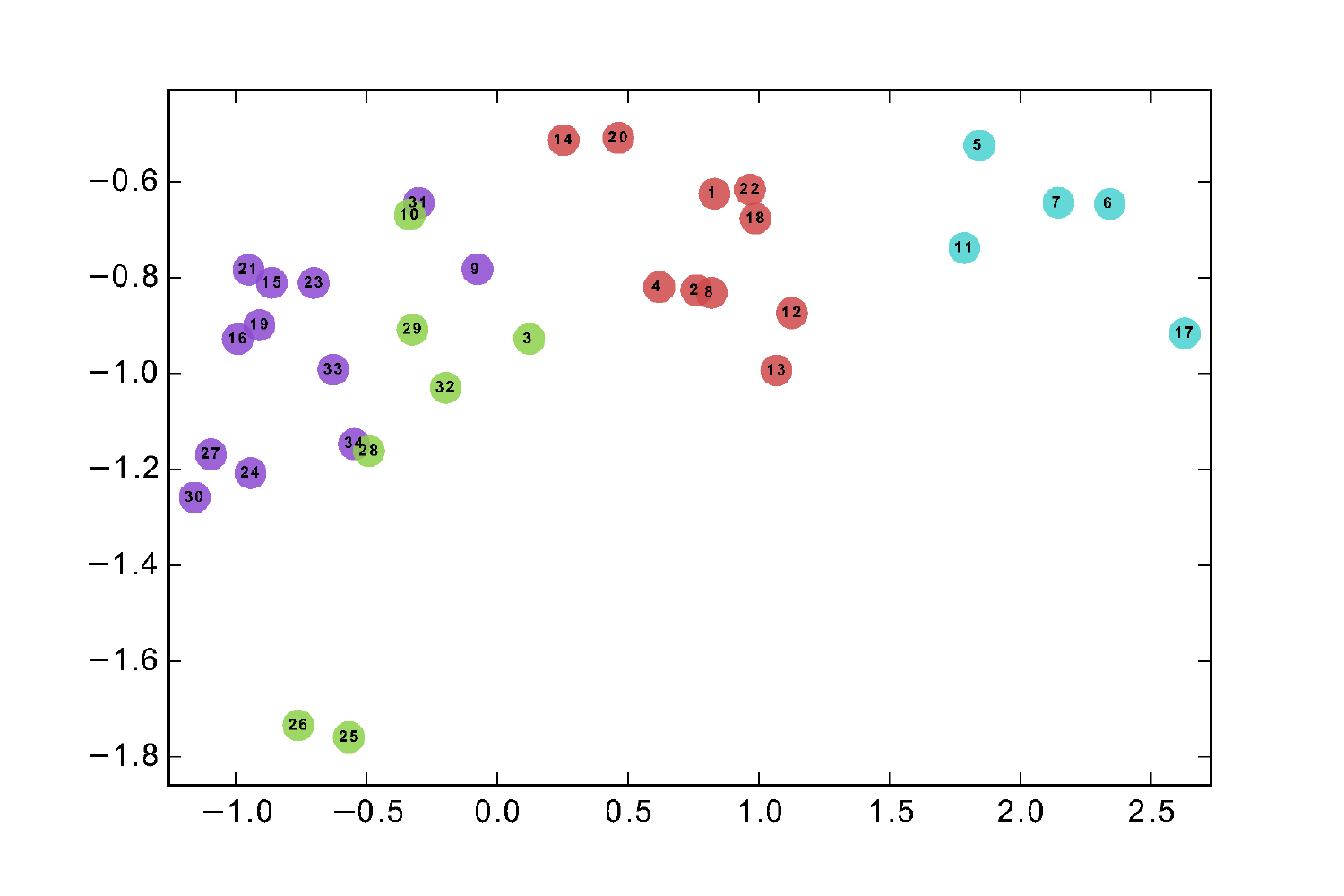}};
\node [anchor=north] at (output.north) {\textbf{Output: }};

\node (input_deepwalk) [textbox, below of=input, yshift=-0.5cm] {DeepWalk: Random walk transition matrix};
\node (graph_sampling_deepwalk) [textbox, below of=graph_sampling, yshift=-0.5cm] {DeepWalk: Truncated random walks};
\node (modeling_deepwalk) [textbox, below of=modeling, yshift=-0.5cm] {DeepWalk: Skip-gram};
\node (output_deepwalk) [textbox, below of=output, yshift=-0.5cm] {DeepWalk: Node Embeddings};

\draw [dashed_arrow] (input) -- (graph_sampling);
\draw [arrow] (graph_sampling) -- (modeling);
\draw [arrow] (modeling) -- (output);
\end{tikzpicture}

\caption{A paradigm for \emph{Deep Learning On Graphs} with DeepWalk's design choice for each building block.}
\label{fig:deep_learning_for_graphs}
\end{figure}
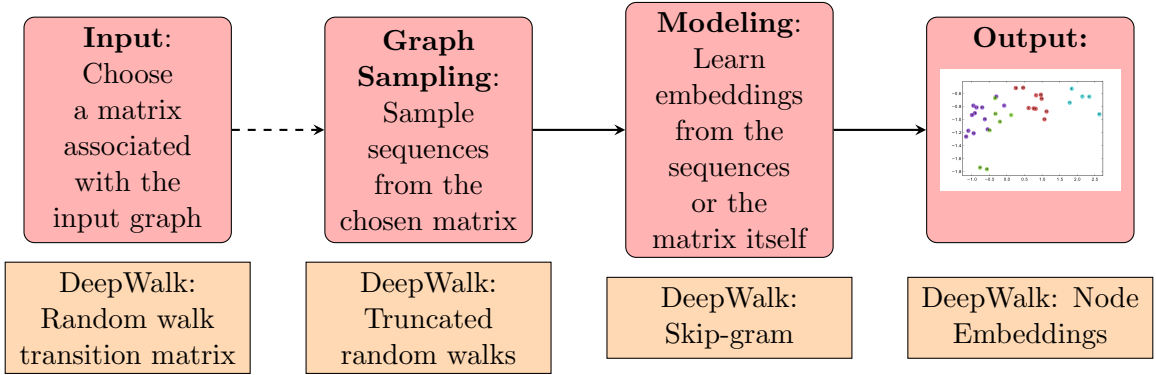

First of all, random walks can be generated on demand. Since the Skip-gram model is also optimized per sample, the combination of random walk and Skip-gram makes DeepWalk an online algorithm.
Secondly, DeepWalk is scalable. Both the process of generating random walks and optimizing the Skip-gram model are efficient and trivially parallelizable.
Most importantly, DeepWalk introduces a paradigm for \emph{deep learning on graphs}, as shown in Figure \ref{fig:deep_learning_for_graphs}.

The first part of the DeepWalk paradigm is choosing a matrix associated with the input graph,
for which DeepWalk chooses the random walk transition matrix.
Indeed, a variety of other choices are also proved to be feasible, such as the normalized Laplacian matrix and the powers of the adjacency matrix.

The second step is graph sampling, where sequences of nodes are implicitly sampled from the chosen matrix.
Note that this step is optional; some network embedding algorithms directly compute the exact matrix elements and build embedding models on it.
However, in a lot of cases graph sampling is a favorable intermediate step for the following two reasons.
First, depends on the matrix of choice, it could take up to quadratic time to compute its exact elements; one example is computing the power series of the adjacency matrix.
In this scenario, graph sampling serves as an scalable approach for approximating the matrix.
Second, compared to a large-scale and sparse graph which is difficult to model,
sequences of symbols are much more easier for deep learning models to deal with.
There are a lot of readily available deep learning methods for sequence modeling, such as RNNs and CNNs.
DeepWalk generates sequence samples via truncated random walk, which effectively extends the neighborhood of graph nodes.

The third step is learning node embeddings from the generated sequences (or the matrix in the first step).
Here, DeepWalk adopts Skip-gram as the model for learning node embeddings,
which is one of the most performant and efficient algorithms for learning word embeddings.

The DeepWalk paradigm is highly flexible, which can be expanded in two possible ways:
\begin{enumerate}
	\item The complexity of the graphs being modeled can be expanded.
	For example, HOPE \cite{ou2016asymmetric} aims at embedding directed graphs, SiNE \cite{sine} and SNE \cite{sne} are methods for embedding signed networks.
	In comparison, the methods of \cite{yang2015network,cene,hsca,gene,wang2017community} are designed for attributed network embedding.
	Much recent work \cite{chang2015heterogeneous,zhao2015representation,li2015learning,hebe,eoe} also attempts to embed heterogeneous networks.
	Besides these unsupervised methods, network embedding algorithms \cite{mmdw,yang2016revisiting,kipf2016semi} have been proposed for semi-supervised learning on graphs. We will have detailed discussion of these methods in Section \ref{sec:attributed_network_embedding} and Section \ref{sec:heterogeneous_network_embedding}.
	\item The complexity of the methods used for the two critical components of DeepWalk, namely sampling sequences from a latent matrix and learning node embeddings from the sampled sequences, can be expanded.
	Much work on network embedding are extensions to DeepWalk's basic framework.
	For instance, \cite{line,node2vec,walklets,grarep} propose new strategies for sequence sampling,
	while \cite{sdne,dngr,grarep} present new stategies for modeling sampled sequences.
	These approachs will be further analyzed in Section \ref{sec:unsupervised_network_embedding}.
\end{enumerate}

\subsection{Notations and Definitions}
Here we introduce the definitions of certain concepts we will used
throughout this survey:
\begin{itemize}
\item
\begin{definition}(graph) \label{def:graph}{\rm
A simple undirected \textbf{graph} $G=(V, E)$ is a collection $V$ of $n$ vertices $v_1, v_2, \cdots, v_n$
together with a set $E$ of edges, which are {\em unordered} pairs of the vertices.
In other words, the edges in an undirected graph have no orientation.

The adjacency matrix $A$ of $G$ is an $n \times n$ matrix where $A_{ij} = 1$ if there is an edge between $v_i$ and $v_j$,
and $A_{ij} = 0$ otherwise.
Unless otherwise stated, we use both graph and network to refer to a simple undirected graph.
}
\end{definition}

\item
\begin{definition}(network embedding) \label{def:network_embedding}{\rm
For a given a network $G$, a \textbf{network embedding} is a mapping function $\Phi: V \mapsto \mathbb{R}^{|V| \times d}$, where $d \ll |V|$.
This mapping $\Phi$ defines the latent representation (or \emph{embedding}) of each node $v \in V$.
Also, we use $\Phi(v)$ to denote the embedding vector for node $v$.
}
\end{definition}

\item
\begin{definition}(directed graph) \label{def:directed_graph}{\rm
A \textbf{directed graph} $G=(V, E)$ is a collection $V$ of $n$ vertices $v_1, v_2, \cdots, v_n$
together with a set $E$ of edges, which are {\em ordered} pairs of the vertices.
The only difference between a directed graph and an undirected graph is that the edges in a directed graph have orientation.
}
\end{definition}

\item
\begin{definition}(heterogeneous network) \label{def:heterogeneous_network}{\rm
A \textbf{heterogeneous network} is a network $G = (V, E)$ with multiple types of nodes or multiple types of edges.
Formally, $G$ is associated with a node type mapping $f_v: v \rightarrow O, \forall v \in V$ and an edge type mapping $f_e: e \rightarrow Q, \forall e \in E$, where $O$ is the set of all node types and $Q$ is the set of all edge types.
}
\end{definition}

\item \begin{definition}(signed graph)
{\rm  A \textbf{signed graph} is a graph where each edge $e \in E$ is associated with a weight $w(e) \in \{-1, 1\}$.
An edge with weight of 1 denotes a positive link between nodes,
whereas an edge with weight of -1 denotes a negative link.  Signed graphs can be used to reflect agreement or trust.
}
\end{definition}

\end{itemize}

\section{Unsupervised Network Embeddings} \label{sec:unsupervised_network_embedding}
In this section, we introduce network embedding methods on simple undirected networks.
We first propose a categorization of the existing methods, and then introduce several
representative methods within each category.

Recent scalable network embedding algorithms are inspired by the emergence of neural language models \cite{bengio2003neural} and {\em word embeddings}, in particular \cite{skipgram, mikolov2013distributed, glove}.
Skip-gram \cite{skipgram} is a highly efficient method for learning word embeddings.
Its key idea is to learn embeddings which are good at predicting nearby words in sentences.
The nearby words $C(w_i)$ (or \emph{context words}) for a certain word $w_i$ in a sentence $w_1, w_2, \cdots, w_T$ are usually defined as
the set of words within a pre-defined window size $k$, namely $w_{i - k}, \cdots, w_{i - 1}, w_{i + 1}, \cdots, w_{i + k}$.
Specifically, the Skip-gram model minimizes the following objective:

\begin{equation}
J = -\sum_{u \in C(w)}log\,Pr(u|w)
\end{equation}
where $Pr(u|w)$ is calculated using a hierarchical or sampled softmax function:

\begin{equation}
Pr(u|w) = \frac{exp(\Phi(w)\cdot\Phi^\prime(u))}{\sum_{u \in W}
{exp(\Phi(w)\cdot\Phi^\prime(u))}}
\end{equation}
Here $\Phi^\prime(u)$ is the distributed representation of $u$ when it serves as a context word, and $W$ is the vocabulary size.

To summarize, the Skip-gram model consists of two phases.
The first phase identifies the context words for each word in each sentence,
while the second phase maximizes the conditional probability of observing the context words given a center word.

By capturing the intrinsic similarity between language modeling and network modeling, DeepWalk \cite{deepwalk} proposed a two-phase algorithm for learning network embedding.
The analogy made by DeepWalk is that nodes in a network can be thought of as words in an artificial language.
Similar to the Skip-gram model for learning word embeddings, the first step of DeepWalk is to identify the \emph{context nodes} for each node.
By generating truncated random walks in the network (which are analogous to sentences),
the context nodes of $v \in V$ can be defined as the set of nodes within a window size $k$ in each random walk sequence,
which can be seen as a combination of nodes from $v$'s $1$-hop, $2$-hop, and up to $k$-hop neighbors.
In other words, DeepWalk learns the network embedding from the combination of $A, A^2, A^3, \cdots, A^k$ where $A^i$ is the $i$-th power of the adjacency matrix.
Once the context nodes have been determined, the second step is same as that of the original Skip-gram model: learn embeddings which maximizes the likelihood of predicting context nodes.
DeepWalk uses the same optimization goal and optimization method as Skip-gram, but any other language model could also be used in principle.

\begin{algorithm}[t]
\begin{algorithmic}[1]
\REQUIRE network $G(V,E)$\\ window size $w$\\ embedding size $d$\\ walks per vertex $\gamma$ \\ walk length $t$
\ENSURE matrix of vertex representations $\Phi \in \mathbb{R}^{|V| \times d}$
	\STATE Initialization: Sample $\Phi$ from $\mathcal{U}^{|V| \times d}$
	\STATE Build a binary Tree $T$ from $V$
	\FOR{$i=0$ to $\gamma$}
		\STATE	$\mathcal{O} = \text{Shuffle}(V)$
        \FOR{\textbf{each} $v_i \in \mathcal{O}$}
		\STATE $\mathcal{W}_{v_i} = RandomWalk(G, v_i, $t$) $
		\STATE SkipGram($\Phi$, $\mathcal{W}_{v_i}$, $w$)
		 \ENDFOR
	\ENDFOR
\end{algorithmic}
\caption{DeepWalk($G$, $w$, $d$, $\gamma$, $t$)}
\label{alg:deepwalk}
\end{algorithm}

Lines 3-9 in Algorithm \ref{alg:deepwalk} shows the core of DeepWalk.
The outer loop specifies the number of times, $\gamma$ of starting random walks at each node.
We can think of each iteration as making a `pass' over the data, sampling one walk per node during this pass.
At the start of each pass, DeepWalk generates a random ordering to traverse the vertices.

In the inner loop, DeepWalk iterates over all the vertices of the graph.
For each node $v_i$ a random walk $|\mathcal{W}_{v_i}| = t$ is generated, and then used to update network embeddings (Line 7).
The Skip-gram algorithm is chosen as the method for updating node representations.

\begin{figure*}[t]
	\centering
	\begin{subfigure}[t]{0.32\textwidth}
		\includegraphics[width=\textwidth]{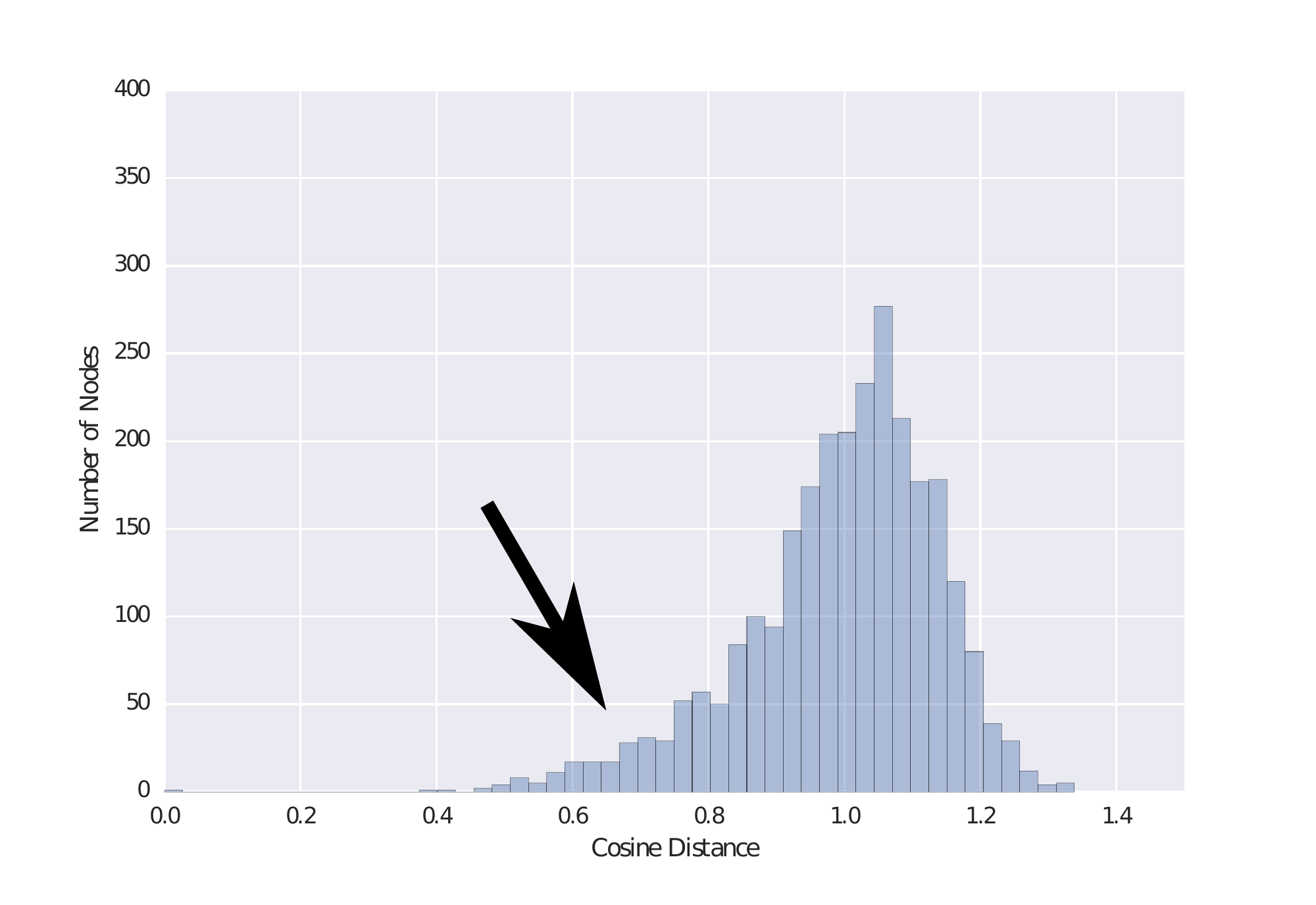}
		\caption{\walklets($A^1$)}
		\label{fig:dist_1}
	\end{subfigure}
	\begin{subfigure}[t]{0.32\textwidth}
		\includegraphics[width=\textwidth]{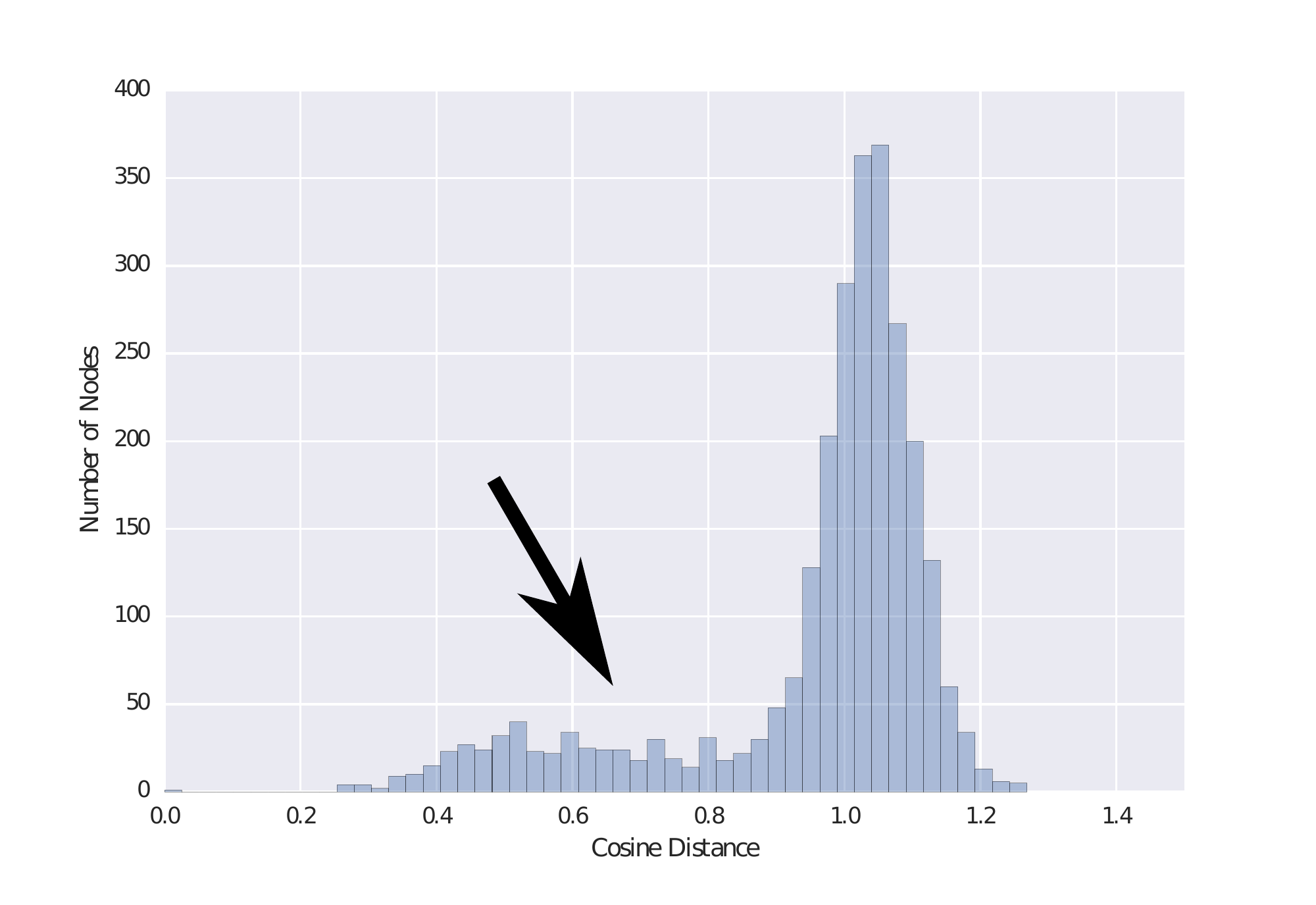}
		\caption{\walklets($A^3$)}
		\label{fig:dist_3}
	\end{subfigure}
	\begin{subfigure}[t]{0.32\textwidth}
		\includegraphics[width=\textwidth]{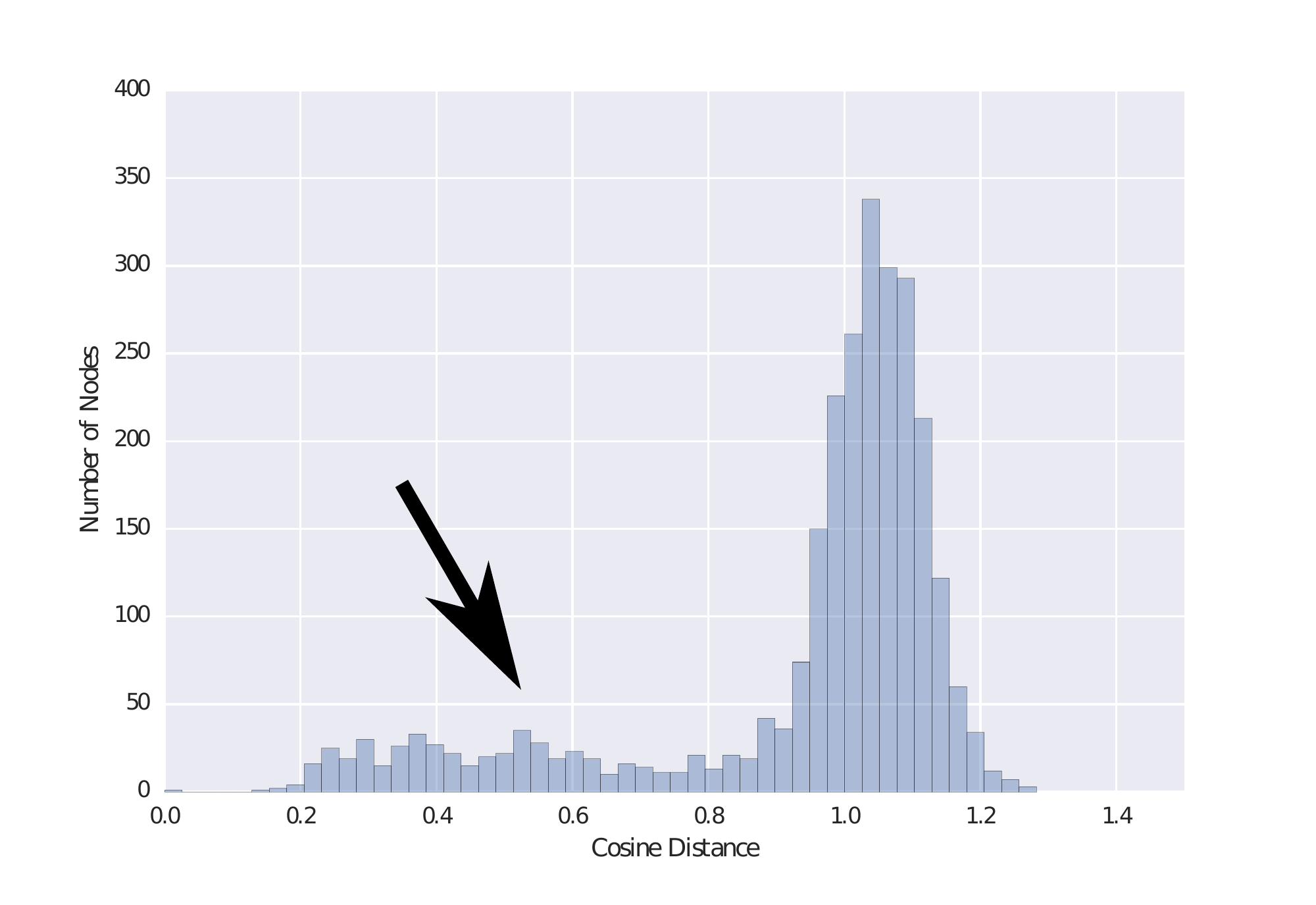}
		\caption{\walklets($A^5$)}
		\label{fig:dist_5}
	\end{subfigure}
\\
	\begin{subfigure}[b]{0.32\textwidth}
		\includegraphics[width=\textwidth]{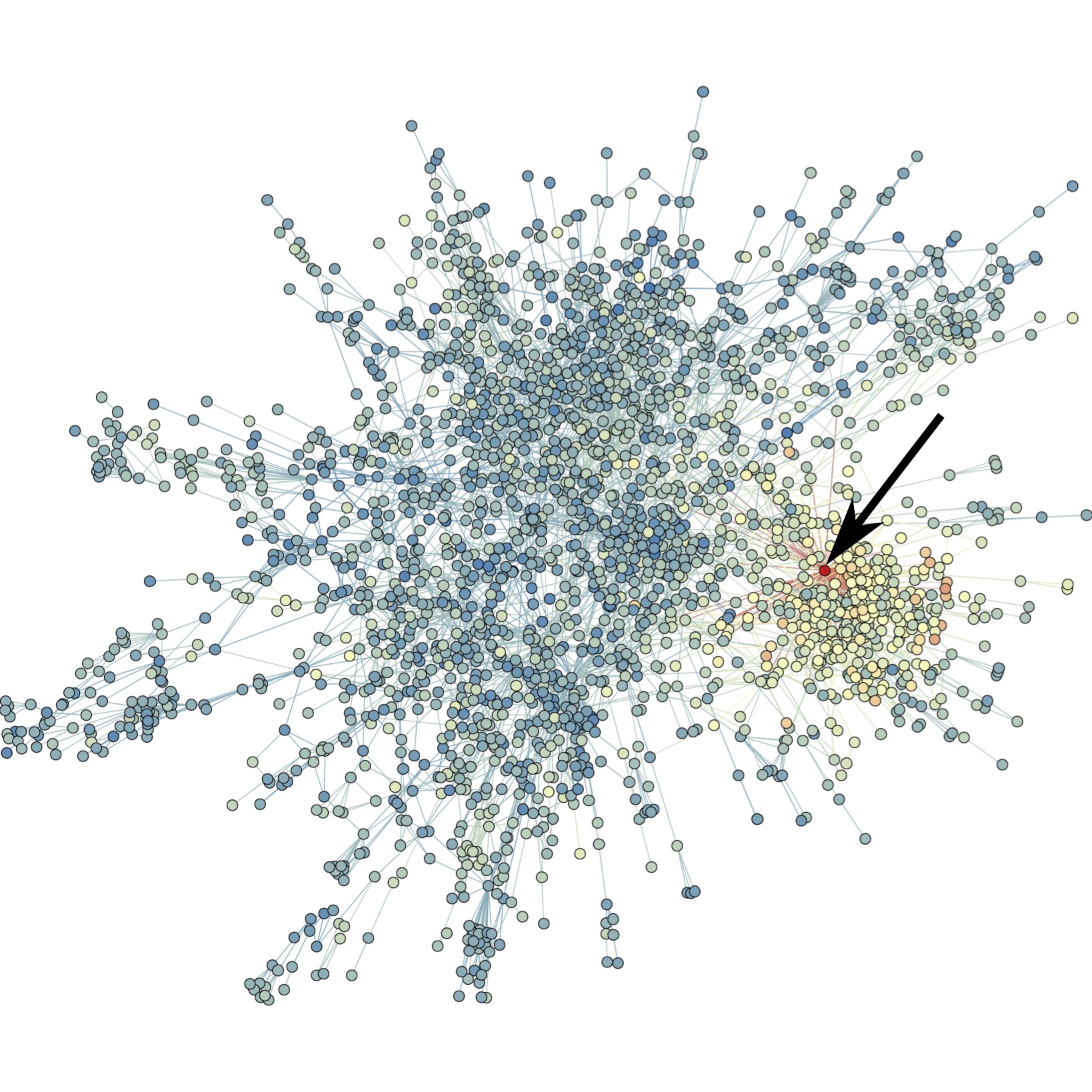}
		\caption{\walklets($A^1$)}
		\label{fig:heatmap_1}
	\end{subfigure}
	\begin{subfigure}[b]{0.32\textwidth}
		\includegraphics[width=\textwidth]{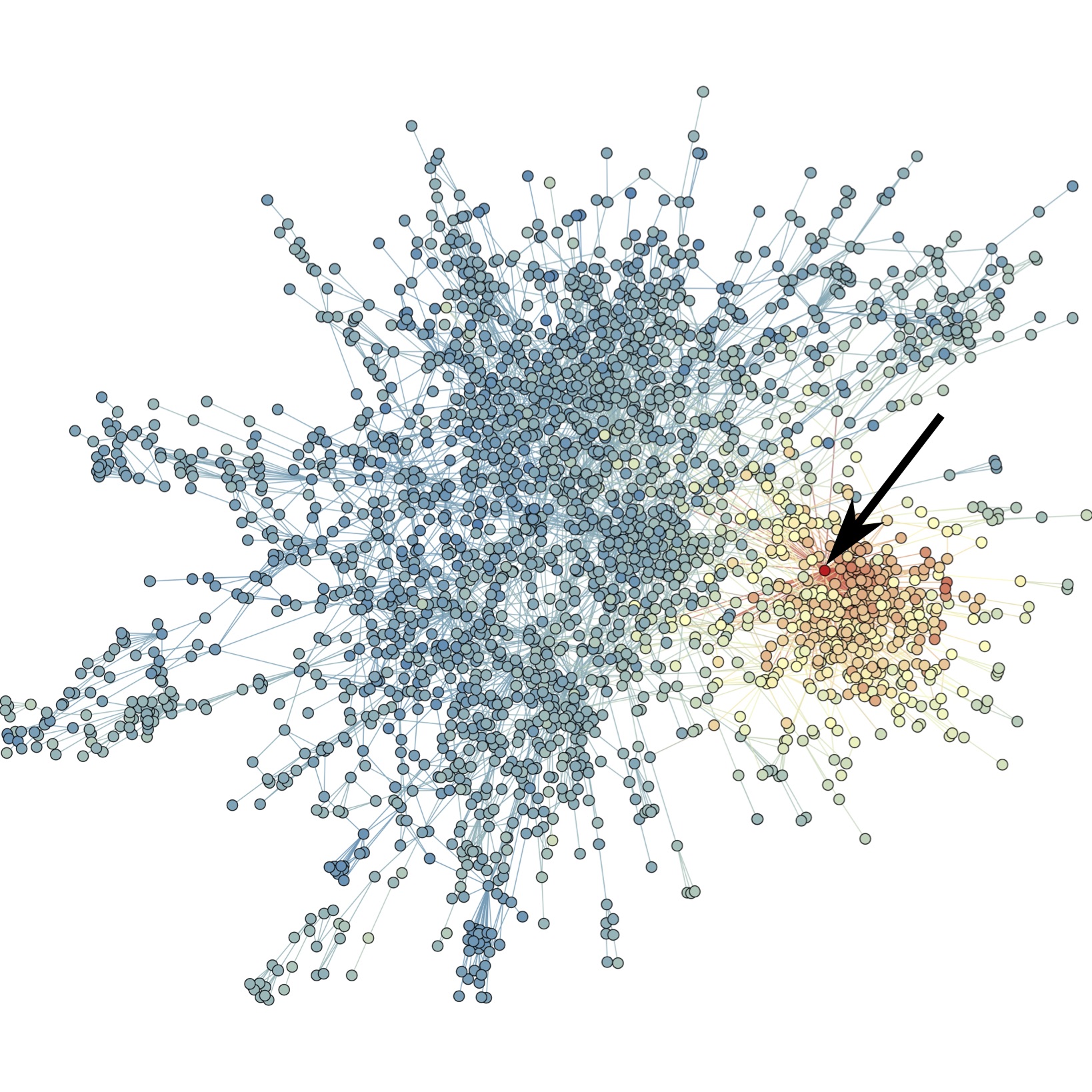}
		\caption{\walklets($A^3$)}
		\label{fig:heatmap_3}
	\end{subfigure}
	\begin{subfigure}[b]{0.32\textwidth}
		\includegraphics[width=\textwidth]{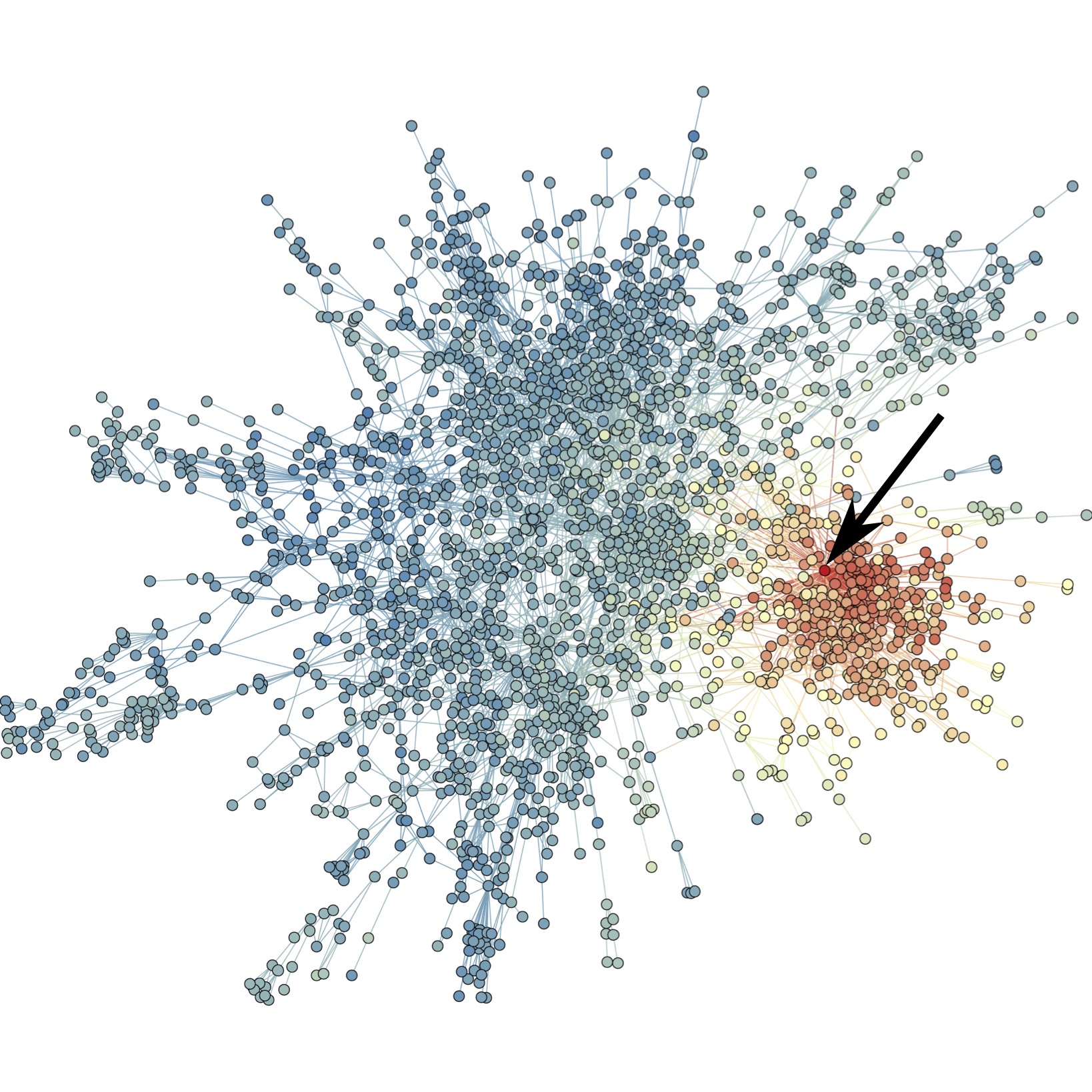}
		\caption{\walklets($A^5$)}
		\label{fig:heatmap_5}
	\end{subfigure}
	\caption{
	Vastly different information can be encoded, depending on the scale of representation chosen.
	Shown in Figures \ref{fig:dist_1}, \ref{fig:dist_3}, and \ref{fig:dist_5} are the distribution of distances to other vertices from $v_{35}$ in the \texttt{Cora} network, at different scales of network representation.
	Coarser representations (such as $A^5$) `flatten' the distribution, making larger communities close to the source vertex.
	Graph heatmap of corresponding distances Figures \ref{fig:heatmap_1}, \ref{fig:heatmap_3}, and \ref{fig:heatmap_5} show the corresponding heatmap of
	cosine distance from vertex $v_{35}$ (shown by arrow) in the \texttt{Cora} network through a series of successively coarser representations.
	Nearby vertices are colored red and distant vertices are colored blue.}
	\label{walklets:fig:multi_scale_visualization_graph}
\end{figure*}

\begin{table}
	\centering
	\begin{tabular}{@{}lll@{}}
		\toprule
		\textbf{Method} & \textbf{Source of Context Nodes} & \textbf{Embedding Learning Method} \\
		\midrule
		DeepWalk \cite{deepwalk} & Truncated Random Walks & Skip-gram with Hierarchical Softmax \\
		LINE \cite{line} & 1-hop and 2-hop Neighbors & Skip-gram with Negative Sampling \\
		Node2vec \cite{node2vec} & Biased Truncated Random Walks & Skip-gram with Negative Sampling \\
		Walklets \cite{walklets} & $A^i$ where $i = 1, 2, \cdots, k$ & Skip-gram with Hierarchical Softmax \\
		GraRep \cite{grarep} & $A^i$ where $i = 1, 2, \cdots, k$ & Matrix Factorization \\
		GraphAttention \cite{graphattention} & $A^i$ where $i = 1, 2, \cdots, k$ & Graph Likelihood \\
		SDNE \cite{sdne} & 1-hop and 2-hop Neighbors & Deep Autoencoder \\
		DNGR \cite{dngr}& Random surfing & Stacked Denoising Autoencoder \\
		\bottomrule
	\end{tabular}
	\caption{Unsupervised network embedding methods categorized by source of context nodes and method for representation learning.}
	\label{tab:network_embedding_methods}
\end{table}

Most subsequent work on graph embeddings has followed this two-phase framework proposed in DeepWalk, with variations in both phases.
Table \ref{tab:network_embedding_methods} summarizes several network embedding methods categorized by different definitions of context nodes
and different methods for learning embeddings:

\begin{itemize}

\item
LINE \cite{line} adopts a breadth-first search strategy for generating context nodes: only nodes which are at most two hops away from a given node are considered as its neighboring nodes.
Besides, it uses negative sampling \cite{mikolov2013distributed} to optimize the Skip-gram model, in contrast to the hierarchical softmax \cite{skipgram} used in DeepWalk.

\item
Node2vec \cite{node2vec} is an extension of DeepWalk which introduces a biased random walking procedure which combines BFS style and DFS style neighborhood exploration.

\item
Walklets \cite{walklets} shows that DeepWalk learns network embeddings from a weighted combination of $A, A^2, \cdots, A^k$.
In particular, DeepWalk is always more biased toward $A^i$ than $A^j$ if $i < j$.
To avoid the above shortcomings, Walklets proposes to learn multiscale network embeddings from each of $A, A^2, \cdots, A^k$.
Since the time complexity of computing $A^i$ is at least quadratic in the number of nodes in the network, Walklets approximates $A^i$ by skipping over nodes in short random walks.
It further learns network embeddings from different powers of $A$ to capture the network's structural information at different granularities.

\item
GraRep \cite{grarep} similarly exploits node co-occurrence information at different scales by raising the graph adjacency matrix to different powers.
Singular value decomposition (SVD) \cite{svd} is applied to the powers of the adjacency matrix to obtain low-dimensional representation of nodes.
There are two major differences between Walklets and GraRep.
First, GraRep computes the exact content of $A^i$, while Walklets approximates it.
Second, GraRep adopts SVD to obtain node embeddings with exact factorization, while Walklets uses the Skip-gram model.
Interestingly, Levy and Goldberg \cite{levy2014neural} proves that skip-gram with negative sampling (SGNS) is implicitly factorizing the PMI matrix between nodes and respective context nodes.
To sum up, GraRep generates network embedding using a process with less noise,
but Walklets proves much more scalable.

\end{itemize}

The models discussed so far rely on some manually chosen parameters to control the distribution of context nodes of each node in the graph.
For DeepWalk, the window size $w$ determines the context node.
Furthermore, the Skip-gram model used has hidden hyper-parameters that determine the importance of an example, based on how far in the context it is.
For Walklets and GraRep, the power to which the graph adjacency matrix is raised to should be decided beforehand.
Selecting these hyperparameters is non-trivial, since they will significantly affect the performance of the network embedding algorithms.

GraphAttention \cite{graphattention} proposes an attention model that learns a multi-scale representation which best predicts links in the original graph.
Instead of pre-determining hyperparameters to control the context nodes distribution, GraphAttention automatically learns the attention over the power-series of the graph transition matrix.
Formally, let $\boldsymbol{D} \in \mathbb{R}^{|V| \times |V|}$ be the co-occurence matrix derived from random walks
and $\tilde{\boldsymbol{P}}^{(0)}$ be the initial random walk starting positions matrix.
GraphAttention parameterizes the expectation of $D$ with a probability distribution $Q = (Q_1, Q_2, \cdots, Q_C)$:
\begin{equation}
	\mathbb{E}[\boldsymbol{D} | Q_1, Q_2, \cdots, Q_C] = \tilde{\boldsymbol{P}}^{(0)}\sum_{k=1}^{C}Q_k(\mathcal{T})^k
\end{equation}
This probability distribution can then be learned by backpropagation from the data itself, e.g.\ by modeling it as the output a softmax layer with parameters ($q_1$, \dots, $q_k$),
\begin{equation}
\mathbb{E}\left. \left[ \mathbf{D}^{\text{softmax}} \ \right| \  q_1,  \dots q_k \right]  \\
= \tilde{\mathbf{P}}^{(0)} \lim_{C \rightarrow \infty} \sum_{j=1}^C \frac{1}{e^{q_j}} \sum_{k=1}^C e^{q_k} \left(\mathcal{T} \right)^k .
\end{equation}
This allows every graph to learn its own distribution Q with a bespoke sparsity and decay form.

The expressiveness of deep learning methods makes them suitable for embedding networks.
SDNE \cite{sdne} learns node representations that preserve the proximity between 2-hop neighbors with a deep autoencoder.
It further preserves the proximity between adjacent nodes by minimizing the Euclidean distance between their representations.
DNGR \cite{dngr} is another deep neural network-based method for learning network embeddings. They adopt a random surfing strategy for capturing graph structural information. They further transform these structural information into a PPMI matrix, and train a stacked denoising autoencoder (SDAE) to embed nodes.

All of these papers focus on embedding simple undirected graphs.
In the next section, we will introduce methods on embedding graphs with different properties, such as
directed graphs and signed graphs.

\subsection{Directed Graph Embeddings}
\label{sec:directed}
The graph embeddings discussed in the previous section were designed to operate on undirected networks.
However, as shown in \cite{zhou2017scalable}, they can be naturally generalized to directed graphs by employing directed random walks as the training data for the network.
Several other recent methods have also been proposed for modeling directed graphs.

HOPE \cite{ou2016asymmetric} is a graph embedding method specifically designed for directed graphs.
HOPE is a general framework for asymmetric transitivity preserving graph embedding, which incorporates several popular proximity measurements such as Katz index, rooted PageRank and common neighbors as special cases. The optimization goal of HOPE is efficiently solved using generalized SVD.

Abu-El-Haija et al. \cite{abu2017learning} propose two separate representations for each node, one where it is a source, and the other where it is a destination.
In this sense, edge embeddings could be thought of as simply a concatenation of the source embedding of the source and the destination of the destination.
These `edge representations' (discussed further in Section \ref{sec:edge_embeddings}),  implicitly preserve the directed nature of the graph.

\subsection{Edge Embeddings}
\label{sec:edge_embeddings}
Tasks like link prediction require accurate modeling of graph edges.
An unsupervised way of constructing a representation for edge $e = (u, v)$ is to apply a binary operator $\circ$ over $\phi(u)$ and $\phi(v)$:
\begin{equation}
	\phi(u, v) = \phi(u) \circ \phi(v)
\end{equation}
In node2vec \cite{node2vec}, several binary operators are considered, such as average, Hardmard product, L1 distance and L2 distance.
However, these symmetric binary operators always assign same representations to edges $(u, v)$ and $(v, u)$,
ignoring the direction of edges.

\begin{figure}[!t]
	\centering
	\includegraphics[width=.6\linewidth]{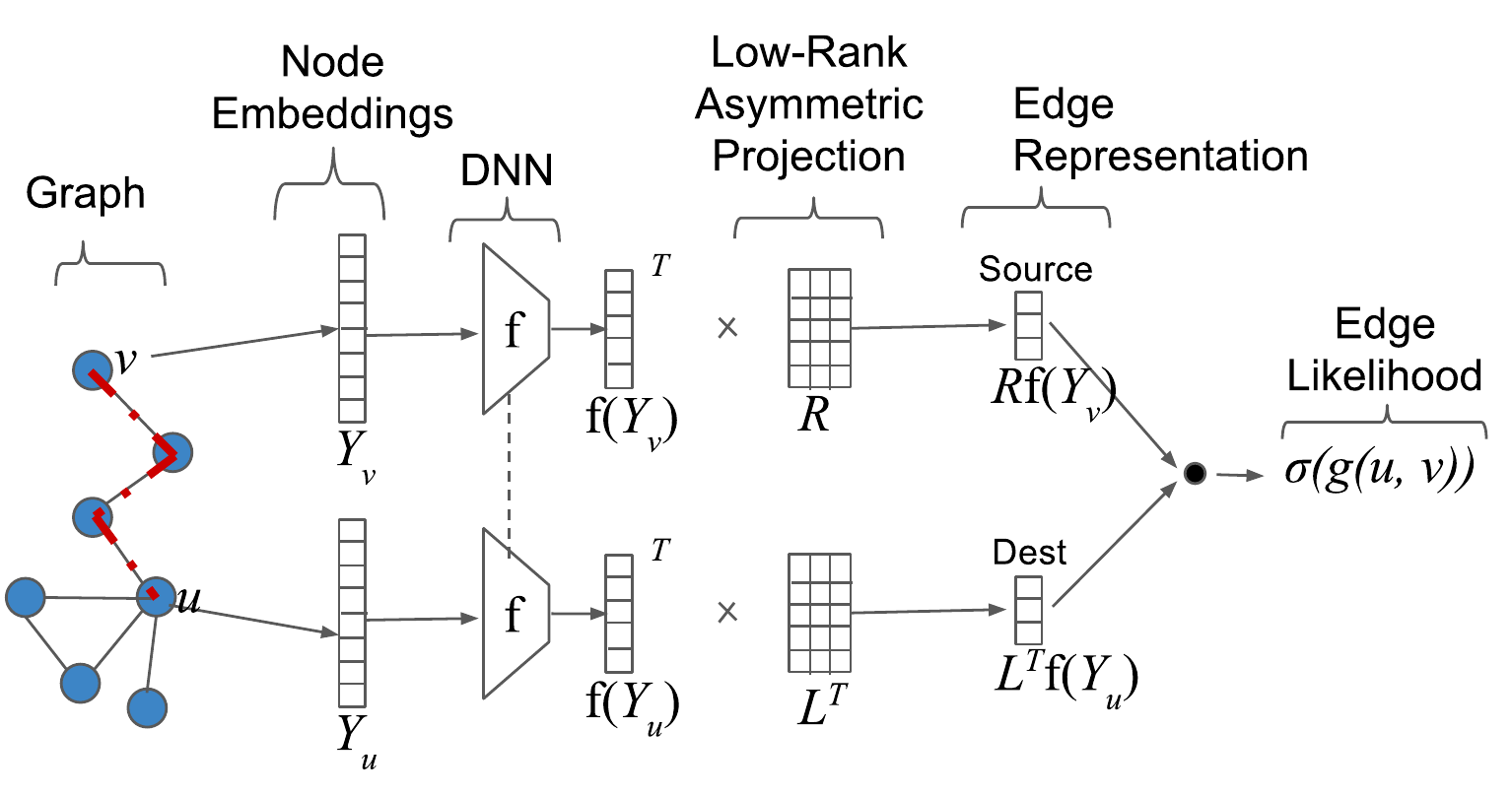}
	\caption{Depiction of the edge representation method in \cite{abu2017learning}.
	On the left: a graph, showing a random walk in dotted-red, where nodes $u, v$ are “close” in the walk (i.e. within a configurable context window parameter).
	Their method accesses the trainable embeddings $Y_u$ and $Y_v$ for the nodes and feed them as input to Deep Neural Network (DNN) $f$.
	The DNN outputs manifold coordinates $f(Y_u)$ and $f(Y_v)$ for nodes $u$ and $v$, respectively.
	A low-rank asymmetric projection transforms $f(Y_u)$ and $f(Y_v)$ to their source and destination representations, which are used by $g$ to represent an edge.
	}
	~\label{fig:edge_representations}
\end{figure}

To alleviate this problem, Abu-El-Haija et al. \cite{abu2017learning} propose to learn edge representations via low-rank asymmetric projections.
Their method consists of three steps.
In the first step, embedding vectors $Y_u \in \mathbb{R}^D$ are learned for every $u \in V$ with node2vec.
Then, a DNN $f_{\theta}: \mathbb{R}^D \rightarrow \mathbb{R}^d$ is learned to reduce the dimensionality of embedding vectors.
Finally, for each node pair $(u, v)$, a low-rank asymmetric projection transforms $f(Y_u)$ and $f(Y_v)$ into their corresponding representations as source and destination nodes, and $\phi(u, v)$ is represented as:
\begin{equation}
	\phi(u, v) = f(Y_u)^T \times M \times f(Y_v)
\end{equation}
where $M$ is the low-rank projection matrix. The model's architecture is further illustrated in Figure \ref{fig:edge_representations}.

\subsection{Signed Graph Embeddings}
Recall that in a signed graph, an edge with weight of 1 denotes a positive link between nodes,
whereas an edge with weight of -1 denotes a negative link.

SiNE \cite{sine} is a deep neural network-based model for learning signed network embeddings.
Based on the structural balance theory, nodes should be closer to their friends (linked with positive edges) than their foes (linked with negative edges).
SiNE preserves this property by maximizing the margin between the embedding similarity of friends and the embedding similarity of foes.
Formally, given a triplet $p = (v_i, v_j, v_k), v_i, v_j, v_k \in V$ where $v_i$ and $v_j$ have a positive link while $v_i$ and $v_k$ have a negative link, the following property holds:
\begin{equation} \label{eq:sine-1}
	f(\Phi(v_i), \Phi(v_j)) \geq f(\Phi(v_i), \Phi(v_k)) + \delta
\end{equation}
where $f$ is a similarity metric between node embeddings and $\delta$ is a tunable margin.
However, negative links are much rarer than positive links in real social networks.
Thus, such a triplet may not exist for many nodes in the network, since there are only positive links in their 2-hop networks.
To solve this problem, an additional virtual node $v_0$ is connected to such nodes with a negative link.
Similarly, given a triplet $p = (v_i, v_j, v_0)$ where a positive link connects $v_i$ and $v_j$ while a negative link connects $v_i$ and $v_0$, we have another objective function:
\begin{equation} \label{eq:sine-2}
	f(\Phi(v_i), \Phi(v_j) \geq f(\Phi(v_i), \Phi(v_0)) + \delta_0)
\end{equation}
The node embeddings are learned by jointly minimizing Eq. \ref{eq:sine-1} and Eq. \ref{eq:sine-2}.

SNE \cite{sne} is a log-bilinear model for signed network embedding.
SNE predicts the representation of a target node by linearly combines the representation of its context nodes.
To capture the signed relationships between nodes, two signed-type vectors are incorporated into the log-bilinear model.

\subsection{Subgraph Embeddings}
Another branch of research concerns embedding larger-scale components of graphs, such as graph sub-structures or whole graphs.
Yanardag and Vishwanathan \cite{yanardag2015deep} present the deep graph kernel, which is a general framework for modeling sub-structure similarity in graphs.
Traditionally, the kernel between two graphs $\mathcal{G}$ and $\mathcal{G^\prime}$ is given by
\begin{equation}
	\mathcal{K(G, G^\prime) = \langle \phi(G), \phi(G^\prime) \rangle_{H}}
\end{equation}
where $\mathcal{\langle \cdot, \cdot \rangle_{H}}$ represents dot product in a RKHS $\mathcal{H}$.

Many sub-structures have been developed to compute this kernel, such as graphlets, subtrees and shortest paths.
However, these representations fail to uncover the similarity between different but similar sub-structures.
That is, even if two graphlets only differ by one edge or one node, they are still considered to be totally different.
This kernel definition causes the \textit{diagonal dominance} problem: a graph is only similar to itself, but not to any other graph.
To overcome this problem, Yanardag and Vishwanathan \cite{yanardag2015deep} present an alternative kernel definition as follows:
\begin{equation}
	\mathcal{K(G, G^\prime) = \phi(G) M \phi(G^\prime)}
\end{equation}
where $\mathcal{M}$ is the similarity matrix between all pairs of sub-structures in the input graph.

To build $\mathcal{M}$, their algorithm first generates the co-occurrence matrix of graph sub-structures.
Then, the Skip-gram model is trained on the co-occurrence matrix to obtain the latent representation of sub-structures,
which is subsequently used to compute $\mathcal{M}$.

\subsection{Meta-strategies for Improving Network Embeddings}
Despite the success of neural methods for network embedding, all methods to date have several shared weaknesses.
Firstly, they are all local approaches -- limited to the structure immediately around a node.
DeepWalk and node2vec adopt short random walks to explore the local neighborhoods of nodes,
while LINE is concerned with even closer relationships (nodes at most two hops away).
This focus on local structure implicitly ignores long-distance global relationships, and the learned representations can fail to uncover important global structural patterns.
Secondly, they all rely on a non-convex optimization goal solved using stochastic gradient descent \cite{skipgram} which can become stuck in a local minima (e.g.\ perhaps as a result of a poor initialization).
In other words, these techniques for learning network embedding can accidentally learn embedding configurations which disregard important structural features of their input graph.

To solve these problems, HARP\cite{harp} proposes a meta strategy for embedding graph datasets which preserves higher-order structural features.
HARP recursively coalesces the nodes and edges in the original graph to get a series of successively smaller graphs with similar structure.
These coalesced graphs, each with a different granularity, provide us a view of the original graph's global structure.
Starting from the most simplified form,
each graph is used to learn a set of initial representations which serve as good initializations for embedding the next, more detailed graph.
This process is repeated until we get an embedding for each node in the original graph.

\begin{figure}[!t]
	\centering
	\begin{subfigure}[b]{.16\linewidth}
		\includegraphics[width=\linewidth]{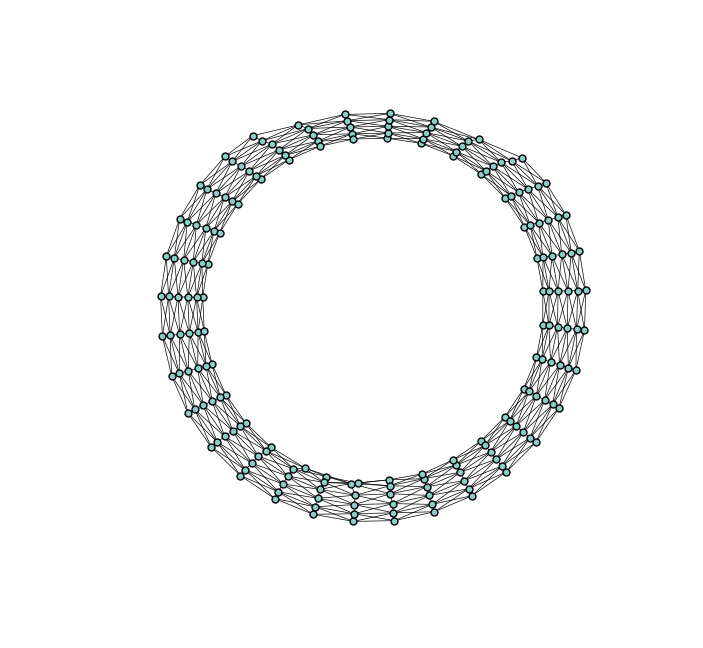}
		\caption{Can\_187}
		\label{fig:can_187_sfdp}
	\end{subfigure}
	\begin{subfigure}[b]{.16\linewidth}
		\includegraphics[width=\linewidth]{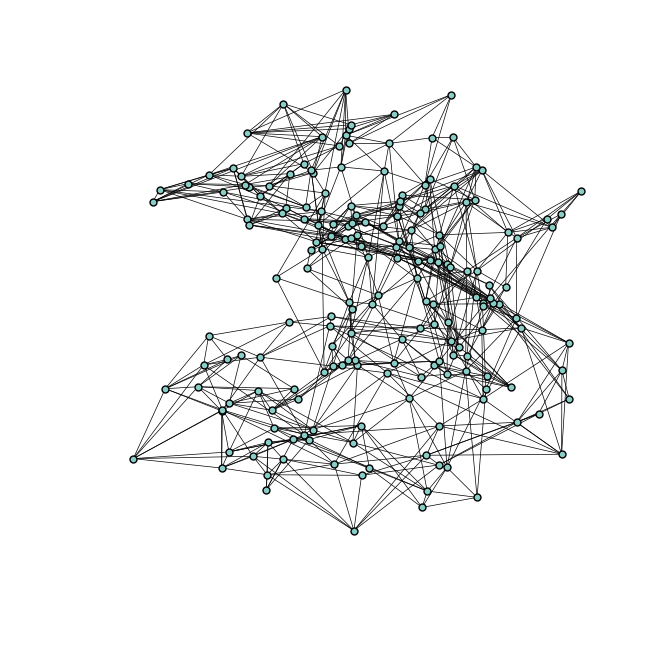}
		\caption{LINE}
		\label{fig:can_187_line}
	\end{subfigure}
	\begin{subfigure}[b]{.16\linewidth}
		\includegraphics[width=\linewidth]{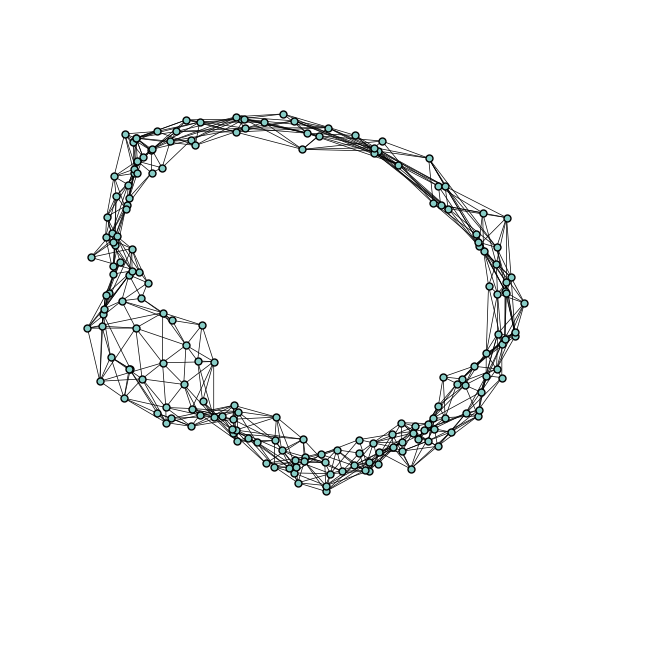}
		\caption{HARP}
		\label{fig:can_187_line_gc}
	\end{subfigure}
	\begin{subfigure}[b]{.16\linewidth}
		\includegraphics[width=\linewidth]{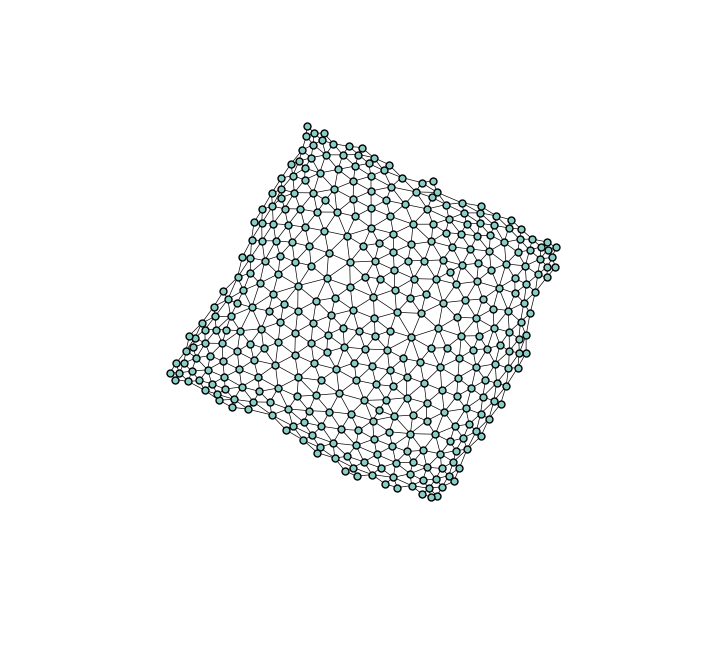}
		\caption{Poisson 2D}\label{fig:poisson_2d_sfdp}
	\end{subfigure}
	\begin{subfigure}[b]{.16\linewidth}
		\includegraphics[width=\linewidth]{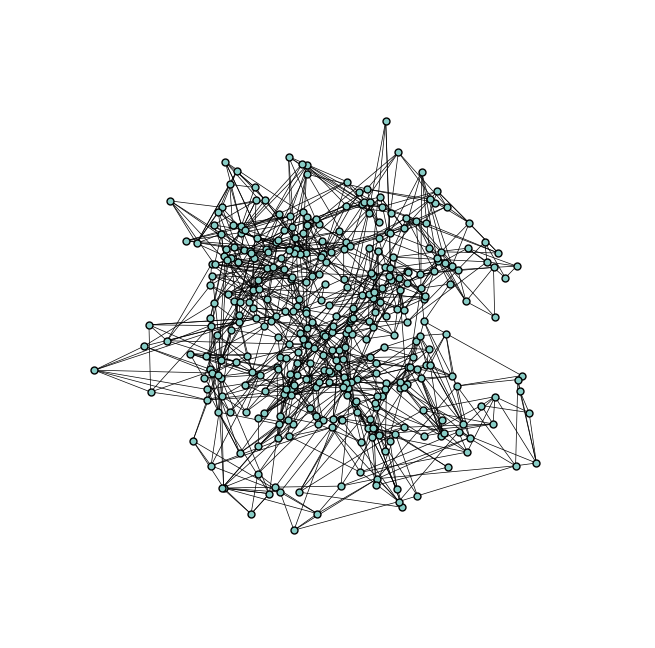}
		\caption{LINE}\label{fig:poisson_2d_line}
	\end{subfigure}
	\begin{subfigure}[b]{.16\linewidth}
		\includegraphics[width=\linewidth]{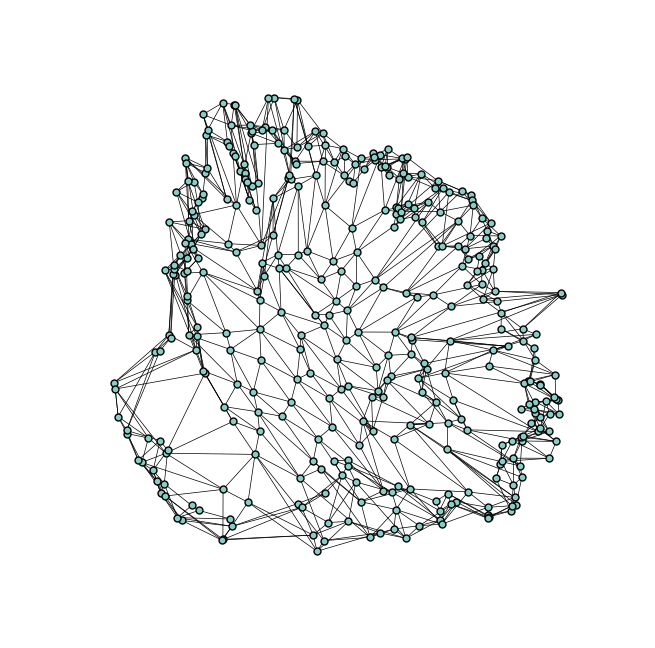}
		\caption{HARP}\label{fig:poisson_2d_line_gc}
	\end{subfigure}
	\caption{Comparison of two-dimensional embeddings from LINE and HARP,
	for two distinct graphs.
	Observe how HARP's embedding better preserves the higher order structure of a ring and a plane.}
	~\label{fig:graph_drawing_comp}
\end{figure}

HARP is a general meta-strategy to improve all of the state-of-the-art neural algorithms for embedding graphs, including DeepWalk, LINE, and Node2vec.
The effectiveness of the HARP paradigm is illustrated in Figure \ref{fig:graph_drawing_comp}, by visualizing the two-dimension embeddings from LINE and the improvement to it, \emph{HARP(LINE)}.
Each of the small graphs we consider have an obvious global structure (that of a ring (\ref{fig:can_187_sfdp}) and a grid (\ref{fig:poisson_2d_sfdp})) which is easily exposed by a force directed layout \cite{hu2005efficient}.
The center figures represent the two-dimensional embedding obtained by LINE for the ring (\ref{fig:can_187_line}) and grid (\ref{fig:poisson_2d_line}).  In these embeddings, the global structure is lost (i.e.\  that is, the ring and plane are unidentifiable).
However, the embeddings produced by using HARP to improve LINE (right) capture both the local and global structure of the given graphs (\ref{fig:can_187_line_gc}, \ref{fig:poisson_2d_line_gc}).

\section{Attributed Network Embeddings} \label{sec:attributed_network_embedding}
The methods we have discussed above leverage only network structural information to obtain network embedding.
However, nodes and edges in real-world networks are often associated with additional features, which are called \emph{attributes}.
For example, in a social network site such as Twitter, the textual contents posted by users (nodes) are available.
Therefore, it is desirable that network embedding methods also learn from the rich content in node attributes and edge attributes.
In the discussion below, we assume that attributes are only associated with nodes, since most existing work focus on exploiting node attributes.
Different strategies have been proposed for different types of attributes.
In particular, researchers are interested in two categories of attributes: high-level features such as text or images, and node labels.

These high-level features are usually high-dimensional sparse features of the nodes, so it is a common practice to use unsupervised text embedding or image embedding models to convert these sparse features into dense embedding features.
Once the embedding features are learned, the major challenge is how to incorporate them into an existing network embedding framework.

TADW \cite{yang2015network} studies the case when nodes are associated with text features.
The authors of TADW first prove that DeepWalk is essentially factorizing a transition probability matrix $M \in \mathbb{R}^{|V| \times |V|}$ into two low-dimensional matrices
$W \in \mathbb{R}^{d\times|V|}$ and $H \in \mathbb{R}^{d\times|V|}$ where $d \ll |V|$.
Inspired by this result, TADW incorporates the text feature matrix $T \in \mathbb{R}^{f_t \times |V|}$
into the matrix factorization process, by factorizing $M$ into the product of $W$, $H$ and $T$.
Finally, $W$ and $H \times T$ are concatenated as the latent representations of nodes.

Another idea is to jointly model network stucture and node features.
Intuitively, in addition to enforcing the embedding similarity between nodes in the same neighborhood,
we should also enforce the embedding similarity between nodes with similar feature vectors.
CENE \cite{cene} is a network embedding method which jointly models network structure and textual content in nodes.
CENE treats text content as a special type of node, and leverages both node-node links and node-content links for node embedding.
The optimization goal is to jointly minimize the loss on both types of links.
HSCA \cite{hsca} is a network embedding method for attributed graphs which models homophily, network topological structure and node features simultaneously.

Besides textual attributes, node labels are another important type of attribute.
In a citation network, the labels associated with papers might be their venue or year of publication.
In a social network, the labels of people cmight be the groups they belong to.
A typical approach to incorporating label information is jointly optimizing the loss for generating node embeddings and for predicting node labels.
GENE \cite{gene} considers the situation when group information is associated with nodes.
GENE follows the idea of DeepWalk, but instead of only predicting context nodes in random walk sequences,
it also predicts the group information of context nodes as a part of the optimization goal.
Wang et al. \cite{wang2017community} present a modularized nonnegative matrix factorization-based method for network embedding which preserves the community structures within network. On the level of nodes, their model preserves first-order and second-order proximities between nodes with matrix factorization; on the level of communities, a modularity constraint term is applied during the matrix factorization process for community detection.

It is also common in real-world networks that node labels are only available for a portion of nodes.
Semi-supervised network embedding methods have been developed for joint learning on both node labels and network structure in such case.
Planetoid \cite{yang2016revisiting} is a semi-supervised network embedding method which learns node representations by jointly predicting the label and the context nodes for each node in the graph.
It works under both inductive and transductive scenarios.
Max-margin DeepWalk (MMDW) \cite{mmdw} is a semi-supervised approach which learns node representations in a partially labeled network.
MMDW consists of two parts: the first part is a node embedding model based on matrix factorization,
while the second part takes in the learned representations as features to train a max-margin SVM classifier on the labeled nodes.
By introducing biased gradients, the parameters in both parts can be updated jointly.

\section{Heterogeneous Network Embeddings} \label{sec:heterogeneous_network_embedding}

Recall that heterogeneous networks have multiple classes of nodes or edges.
To model the nodes and edges of different types, most network embedding methods we introduce below learn node embeddings via jointly minimizing the loss over each modality.
These methods either directly learn all node embeddings in the same latent space, or construct the embeddings for each modality beforehand and then map them to the same latent space.

Chang et al. \cite{chang2015heterogeneous} present a deep embedding framework for heterogeneous networks.
Their model first constructs a feature representation for each modality (such as image, text), then maps the embeddings of different modalities into the same embedding space.
The optimization goal is to maximize the similarity between the embeddings of linked nodes, while minimizing that of the unlinked nodes.
Note that edges can be between both nodes within the same modality as well as nodes from different modalities.

Zhao et al. \cite{zhao2015representation} is another such framework for constructing node representations in a heterogeneous network. Specifically, they consider the Wikipedia network with three types of nodes: entities, words and categories.
The co-occurrence matrices between same and different types of nodes are built, and the representations for entities, words and categories are jointly learned from all matrices using coordinate matrix factorization.

Li et al. \cite{li2015learning} propose a neural network model for learning user representations in a heterogeneous social network. Their method jointly models user-generated texts, user networks and multifaceted relationships between users and user attributes.

HEBE \cite{hebe} is an algorithm for embedding large-scale heterogeneous event networks, where an event is defined as the interaction between a set of nodes (possibly of different types) in the network.
While previous work decomposes an event into the pairwise interaction between each pair of nodes involved in the event,
HEBE treat the whole event as a hyperedge and preserves the proximity between all participating nodes simultaneously.
Specifically, for each node in a hyperedge, HEBE considers it as the target node and the remaining nodes in the hyperedge as context nodes.
Thus, the underlying optimization goal is to predict the target node given all context nodes.

EOE \cite{eoe} is a network embedding method for coupled heterogeneous networks, where two homogeneous networks are connected with inter-network edges.
EOE learns latent node representations for both networks, and utilize a harmonious embedding matrix to transform the representations of different networks into the same space.

Besides modeling heterogeneous nodes and edges jointly, another promising direction of work is on extending random walks and embedding learning methods to a heterogeneous scenario.
Metapath2vec \cite{dong2017metapath2vec} is an extension to DeepWalk which works for heterogeneous networks.
For constructing random walks, metapath2vec uses meta-path-based walks which capture the relationship between different types of nodes.
For learning representation from random walk sequences, they propose heterogeneous Skip-gram which considers node type information during model optimization.

\section{Applications of Network Embeddings}

Network embeddings have been widely employed in practice, due to their ease of use in turning adjacency data into actionable features.
Here we review several representative applications of network embeddings to demonstrate how they can be used:

\subsection{Knowledge Representation}
The problem of knowledge representation is concerned with encoding facts about the world using short sentences (or tuples) composed of subjects, predicates, and objects.
While it can be viewed as strictly as a heterogeneous network, it is an important enough application area to mention here in its own right:

\begin{itemize}
\item
GenVector \cite{yang2015multi} studies the problem of learning social knowledge graphs, where the goal is to connect online social networks to knowledge bases.
Their multi-modal Bayesian embedding model utilizes DeepWalk for generating user representations in social networks.

\item
RDF2Vec \cite{ristoski2016rdf2vec} is an approach for learning latent entity representations in Resource Description Framework (RDF) graphs.
RDF2Vec first converts RDF graphs into sequences of graph random walks and Weisfeiler-Lehman graph kernels,
and then adopt CBOW and Skip-gram models on the sequences to build entity representations.

\end{itemize}

\subsection{Recommender Systems}
Another branch of work attempts to incorporate network embeddings into recommender systems. Naturally, the interactions between users, users' queries and items altogether form a heterogeneous network which encodes the latent preferences of users over items. Network embedding on such interaction graphs could serve as an enhancement to recommender systems.

\begin{itemize}

\item
Chen et al. \cite{chen2015exploiting} exploit the usage of social listening graph to enhance music recommendation models.
They utilize DeepWalk to learn latent node representations in the social listening graph, and incorporate these latent representations into factorization machines.

\item
Chen et al. \cite{chen2016query} propose Heterogeneous Preference Embedding to embed user preference and query intention into low-dimensional vector space. With both user preference embedding and query embedding available, recommendations can be made based on the similarity between items and queries.

\end{itemize}

\subsection{Natural Language Processing}
State-of-the-art network embedding methods are mostly inspired by advances in the field of natural language processing, especially neural language models. At the same time, network embedding methods also lead to better modeling of human language.

\begin{itemize}
\item
PLE \cite{ren2016label} studies the problem of label noise reduction in entity typing. Their model jointly learns the representations of entity mentions, text features and entity types in the same feature space. These representations are further used to estimate the type-path for each training example.
\item
CANE \cite{tu2017cane} is a context-aware network embedding framework.
They argue that one node may exhibit different properties when interacting with different neighbors, thus its embedding with respect to these neighbors should be different.
CANE achieves this goal by employing mutual attention mechanism.
\item
Fang et al. \cite{fang2016community} propose a community-based question answering (cQA) framework which leverages the social interactions in the community for better question-answering matching. Their framework treats users, questions and answers and the interactions between them as a heterogeneous network and trains a deep neural network on random walks in the network.
\item
Zhao et al. \cite{zhao2016expert} study the problem of expert finding in community-based question answering (cQA) site.
Their method adopts the random-walk method in DeepWalk for embedding social relations between users and RNNs for modeling users' relative quality rank to questions.
\end{itemize}

\subsection{Social Network Analysis}
Social networks are prevailing in the real world, and it is not suprising that network embedding methods have become popular in social network analysis.
Network embeddings on social network have prove to be powerful features for a wide spectrum of applications, leading to improved performance on a lot of downstream tasks.

\begin{itemize}
\item
Perozzi et al. \cite{perozzi2015exact} study the problem of predicting the exact age of users in social networks.
They learn the user representations in social networks with DeepWalk, and adopts linear regression on these user representations for age prediction.

\item
Yang et al. \cite{yang2016neural} propose a neural network model for modeling social networks and mobile trajectories simultaneously.
They adopt DeepWalk to generate node embeddings in social networks and the RNN and
GRU models for generating mobile trajectories.

\item
Dallmann et al. \cite{dallmann2016extracting} show that by learning Wikipedia page representations from both the Wikipedia link network and Wikipedia click stream network with DeepWalk,
they can obtain concept embeddings of higher quality compared to counting-based methods on the Wikipedia networks.

\item
Liu et al. \cite{liu2016aligning} propose Input-output Network Embedding (IONE), which use network embeddings to align users across different social networks. IONE achieves this by preserving the proximity of users with similar followers and followees in a common embedding space.

\item
Chen and Skiena \cite{chen2017vector} demonstrate the efficacy of network embedding methods in measuring similarity between historical figures.
They construct a network between historical figures from the interlinks between their Wikipedia pages, and use DeepWalk  to obtain vector representations of historical figures.
It is shown that the similarity between the DeepWalk representations of historical figures can be used as an effctive decent similarity measurement.

\item DeepBrowse \cite{deepbrowse} is an approach for browsing through large lists in the absence of a predefined hierarchy.
DeepBrowse is defined by the interaction of two fixed, globally-defined permutations on the space of objects: one ordering the items by similarity, the second based on magnitude or importance.
The similarity between items is computed by using DeepWalk embeddings generated over the interaction graph of objects.

\item
TransNet \cite{tu2017transnet} is a translation-based network embedding model which exploits the rich semantic information in graph edges for relation prediction on edges.
TransNet treats the interactions between nodes as a translation operation and further employ a deep autoencoder to construct edge representations.

\end{itemize}

\subsection{Other Applications}
\begin{itemize}
\item
Geng et al. \cite{geng2015learning} and Zhang et al. \cite{zhang2016learning} develop deep neural network models which learns distributed representations of both users and images from an user-image co-occurrence network.
The representation learning process in the network is analogous to that of DeepWalk \cite{deepwalk}, except that they also incorporate image features extracted with a DCNN into the optimization process.
\item Wu et al. \cite{wu2016learning} treat the click data collected from users' searching behavior in image search engines as a heterogeneous graph.
The nodes in the click graph are text queries and images returned as search results, while the edges indicates the click count of an image given a search query.
By proposing a neural network model based on truncated random walks, their method learns multimodal representations of text and images,
which are shown to boost cross-modal retrieval performance on unseen queries or images.
\item Zhang et al. \cite{zhang2015learning} apply DeepWalk to large-scale social image-tag collections to learn both image features and word features in a unified embedding space.
\end{itemize}

These applications only represent the tip of the iceberg.
The future of network embeddings seems bright, with new algorithmic approaches
producing better embeddings to feed increasingly sophisticated neural networks
as Deep Learning continues to grow in popularity and importance.

\section{Conclusions and Future Directions}
Network embedding is an exciting and rapidly growing research area which attracts researchers from various communities, especially data mining, machine learning and natural language processing.
While most work concerned about general methods for network embedding, we argue that the applications of network embedding is even more underresearched.
We anticipate a large body of work on additional applications of network embeddings, such as improving the performance of natural language processing and information retrieval models,
mining biology network and social networks, to name a few.

Also, much work has been done for graphs which possess different properties and from different domains.
In terms of graph properties, various methods are proposed for directed graphs, signed graphs, heterogeneous graphs and attributed graphs.
In terms of application domains, network embedding methods are applied to a wide spectrum of graphs including knowledge graphs, biology graphs and social networks.
However, doubtlessly much more work can be done on this front by exploiting the unique characteristics of these graphs.

\subsection{The search for the right context}
Inspired by the two-phase network embedding learning framework presented in DeepWalk,
various strategies have been proposed for searching for the right context, as discussed in Table \ref{tab:network_embedding_methods}.
However, most of these strategies relies on a rigid definition of context nodes identical for all networks, which is not desirable.

Under this background, there is much effort recently on unifying different network embedding under a general framework \cite{chen2017fast,qiu2017network}.
GEM-D \cite{chen2017fast} decomposes graph embedding algorithms into three building blocks: node proximity function, warping function and loss function.
They show that algorithms such as Laplacian Eigenvectors, DeepWalk, LINE, and node2vec can all be unified under this framework.
By testing different design choices for each building block on real-world graphs, they pick the triple which works the best empirically:
the combination of the finite-step transition matrix, exponential warping function and warped Frobenius norm loss.
However, such design decisions are purely made based on models' empirical performance on a limited number of networks, which may not work well for all networks.

A promising approach is the attention model recently proposed in GraphAttention\cite{graphattention}.
By parameterizing the attention over the power series of the transition matrix, GraphAttention automatically learns different attention parameters for different networks.
\subsection{Improved Losses / Optimization Models}
Another issue with the neural embedding methods is their dependence upon general loss functions and optimization models, such as Skip-gram.
These optimization goals and models are not tuned for any particular task.
As a result, though the learned network embeddings have been proven to achieve competitive performance on a variety of tasks such as node classification and link prediction,
they are suboptimal when compared with end-to-end embeddings methods designed specifically for a task.

Thus, another future direction for network embedding algorithms is to design loss functions and optimization models for a specific task.
From this perspective, the semi-supervised network embedding methods can been seen as specifically designed for the node classification task.
Another attempt is made by Abu-El-Haija et al. \cite{abu2017learning}, where the \textbf{graph likelihood} is proposed as a novel objective tuned for link prediction.
Given a training graph $G = (V, E_{train})$, its graph likelihood is defined as a product of an edge estimate $Q$ over all node pairs:
\begin{equation}
	Pr(G) = \prod_{(u, v) \in E_{train}}Q(u, v) \prod_{(u, v) \notin E_{train}}1 - Q(u, v)
\end{equation}
where $Q: V \times V \rightarrow [0, 1]$ is a trainable edge estimator.

\bibliographystyle{plain}
\bibliography{bibtex_example}
\end{document}